\documentclass[10pt]{article}

\usepackage{amsmath}
\usepackage{amssymb}

\usepackage{graphicx}

\usepackage{cite}

\usepackage{color} 

\usepackage[labelfont=bf,labelsep=period,justification=raggedright]{caption}

\makeatletter
\renewcommand{\@biblabel}[1]{\quad#1.}
\makeatother

\date{}

\begin{document}

\begin{flushleft}
{\Large
\textbf{The frustrated brain: From dynamics on motifs to communities and networks}
}
\\
\medskip 

Leonardo L. Gollo$^{1,\ast}$,  
Michael Breakspear$^{1,2,3}$
\\
\medskip

\bf{1} Systems Neuroscience Group, QIMR Berghofer Medical Research Institute, Brisbane, Queensland, Australia
\\
\bf{2} School of Psychiatry, University of New South Wales and The Black Dog Institute, Sydney, New South Wales, Australia
\\
\bf{3} The Royal Brisbane and Women's Hospital, Brisbane, Queensland, Australia
\\
\smallskip
$\ast$ E-mail: leonardo.l.gollo@gmail.com
\end{flushleft}

\section*{Abstract}

%
Cognitive function depends on an adaptive balance between flexible dynamics and integrative processes in distributed cortical networks. Patterns of zero-lag synchrony likely underpin numerous perceptual and cognitive functions. Synchronization fulfils integration by reducing entropy, whilst adaptive function mandates that a broad variety of stable states be readily accessible. Here, we elucidate two complementary influences on patterns of zero-lag synchrony that derive from basic properties of brain networks. First, mutually coupled pairs of neuronal subsystems -- resonance pairs -- promote stable zero-lag synchrony amongst the small motifs in which they are embedded, and whose effects can propagate along connected chains. Second, frustrated closed-loop motifs disrupt synchronous dynamics, enabling metastable configurations of zero-lag synchrony to coexist. We document these two complementary influences in small motifs and illustrate how these effects underpin stable versus  metastable phase-synchronization patterns in prototypical modular networks and in large-scale cortical networks of the macaque (CoCoMac). We find that the variability of synchronization patterns depends on the inter-node time delay, increases with the network size, and is maximized for intermediate coupling strengths. We hypothesize that the dialectic influences of resonance versus frustration may form a dynamic substrate for flexible neuronal integration, an essential platform across diverse cognitive processes. 

\vspace{0.3 cm}
{\bf Keywords:} functional network, dynamic functional connectivity, neural mass model, anti-phase synchronization, macaque cortical network


\section*{Introduction}

Understanding large-scale cortical dynamics and their relationship to the underlying anatomy is key to unlocking the computational principles of the brain. Despite cytological differences at the microscopic level, the repetitive design of mesoscopic motifs -- circuits, columns and local hierarchies -- is a stand-out feature of brain anatomy. By promising a way of scaling up in size, this repetitive principle is likely crucial to the success of the very large-scale computational modeling efforts currently underway. From this view, it is the larger scale structural connectivity and topological constraints of macroscopic anatomy that plays the crucial role in sculpting information processing in the cortex~\cite{Passingham02,Sporns10}. According to this modern connectionist approach, the anatomical network is fundamental because the functionally specialized and integrative roles of cortical regions come from their interactions with other brain regions, not only from their microscopic particularities. Functional 
network is a widely used approach to study one of the major current paradigms of neuroscience: the relationship between structure and dynamics~\cite{Honey10}. Whilst the physical structure remains roughly constant on the time scale of minutes to hours, the dynamic states of cortical regions evolve in a coordinated way to perform a multitude of tasks, which vary on far shorter time scales. Although some states of brain dynamics are very general and common, at least among individuals of a same species, others are very particular, and perhaps unique. 
 
Spanning different neuronal scales, synchronization is an important dynamical feature that can increase the effectiveness of interactions between brain regions, for example by aligning neuronal spikes from different regions into critical windows for spike-time dependent plasticity~\cite{Masquelier09}. Indeed, an emerging paradigm considers synchronization as \emph{the} key feature that modulates cortical interactions~\cite{Fries05}: Accordingly, its maximum effectiveness is achieved when two mutually interacting areas are phase coupled. The synchronization strength between regions is thus fundamental, and a functional network is composed of nodes that are linked by highly correlated pairs. The interactions between brain regions, in turn, are not static but rather erratic, depending on their dynamics and past activity~\cite{Breakspear04,Hutchison13,Zalesky14}. Such a dynamic balance of integration and instability likely underlies the need for cortical function to switch according to the changing environmental 
needs~\cite{Friston00,Breakspear02b,Friston12}. Hence, variable and flexible dynamics, which depends on coupling delay~\cite{Kutchko13} and the balance of excitation and inhibition~\cite{Yang12}, is a hallmark of adaptive cognitive behaviour.

Understanding dynamic functional networks is not only an important topic, but also a rather complex one. As with any problem endowed with complexity, a wise strategic approach is required. A standard way to reduce the level of complexity is to divide the problem into pieces. Often, however, it is not clear which is the best way to split the problem. Neuronal systems are clearly more than the sum of their parts: Extrapolations thus demand careful analysis.  The fingerprints of small motifs have already become a standard procedure to understand the structure of complex networks~\cite{Milo02, Sporns04, Song05, Sporns07, Rubinov10, Harriger12}.  A recent, promising approach to understanding the relationship between structure/dynamics on complex networks consists of studying the structure/dynamics relationship in network motifs~\cite{Eguiluz11, Zhao11, Trousdale12, Gollo14}. This rests on the premise that the dynamical features of network motifs plays a role in shaping the dynamics and synchronization of complex 
networks~\cite{Zhao11, Trousdale12}.

The connections between cortical regions in the brain, as derived from the primate cortical network obey a few general principles that may inform the study of structure/dynamic analyses~\cite{Bullmore09}: First, the brain topology is hierarchical with cortical regions occupying the highest positions in the hierarchy~\cite{Felleman91,Breakspear05,Stam12}.  Second, spatial clustering leads to a modular architecture, in which nodes are strongly interconnected with regions within the same cluster -- typically nearby in space -- than with regions outside their cluster~\cite{Hilgetag04,Zemanova06}. 
Third, some nodes can be classified as hubs because they have a degree larger than the average node degree of the network~\cite{Hagmann08, Zamora10, van13, Sporns07}. These principles have been linked to evolutionary and functional advantages as well as optimization arising from the physical and metabolic constraints under which brains operate on evolutionary time scales~\cite{Bullmore12,Chen13}. These general network properties can also shape the dynamics and synchronization of cortical networks~\cite{Zhou06,Honey07,Zhou07}. 
However, a complete understanding of how the structural constraints of the anatomical connectivity influence dynamics is lacking, particularly regarding the mechanisms underlying the variability of synchronization. 

Here we analyse the synchronization patterns as they arise in small motifs, prototypical modular structures, and cortical networks. We aim to identify factors that enrich the landscape of synchronized cortical states. The paper is structured as follows: We first describe mesoscopic dynamics as they arise from small motifs of coupled neural masses. We retrace recent work that highlights the stablising role of reciprocally coupled pairs (which we call resonant pairs) on the motifs (small subnetworks of 3 nodes) in which they are embedded. This influence is contrasted with that of closed loops, which introduce frustration into the system. Frustrated motifs hence disrupt stability and synchrony, introducing metastability and variability. We then study these two competing influences in small constructed modular networks. We subsequently examine structural networks derived from tracing studies of the Macaque brain, using the distribution of motifs to link the observed dynamics to those on our prototypical 
networks.

\section*{Neuronal network dynamics}

We employed numerical methods to study neuronal dynamics on large-scale cortical networks. Each node was considered as a conductance based neural mass model~\cite{Breakspear02, Honey07, Honey09}, described in detail in the appendix. This approach reduces the dimensionality of a cortical region in each node to three nonlinear differential equations, which govern the dynamics of the excitatory subpopulation, the inhibitory subpopulation and the fraction of open potassium channels. Analysis of such a reduced model is required to render the problem of large-scale cortical dynamics tractable. Distributed cortical regions are coupled together through long-range excitatory connections with a delay of 10 ms (unless otherwise stated) to account for finite axonal conduction speed~\cite{Tomasi12}. In our mesoscale model, the dynamic variables start from random initial conditions. Exploring the phase space can lead to different basins of attraction and different transient dynamics, causing considerable trial-to-trial variability, consistent with experiments~\cite{Arieli96,Poulet08,Cafaro10}. This variability is an intrinsic feature of many large-dimensional systems. Occasionally, such variability can also arise from transitions between metastable states. The following results should not be interpreted as deterministic outcomes of the model, but instead as likely statistical outcomes revealed through a large number of repetitions of the same numerical experiment. Deeper statistical features, in contrast, are remarkably persistent. For instance, nodes oscillate with two dominant frequencies, a slow $\sim$10 Hz and a fast $\sim$100 Hz time scale, and two mutually coupled nodes typically synchronize in anti-phase synchrony~\cite{Gollo14}.

\subsection*{Resonance pairs in frustrated motifs promote variable synchronization patterns}

\begin{figure}[!h]
\begin{center}
\includegraphics[angle=0,width=1.05\columnwidth]{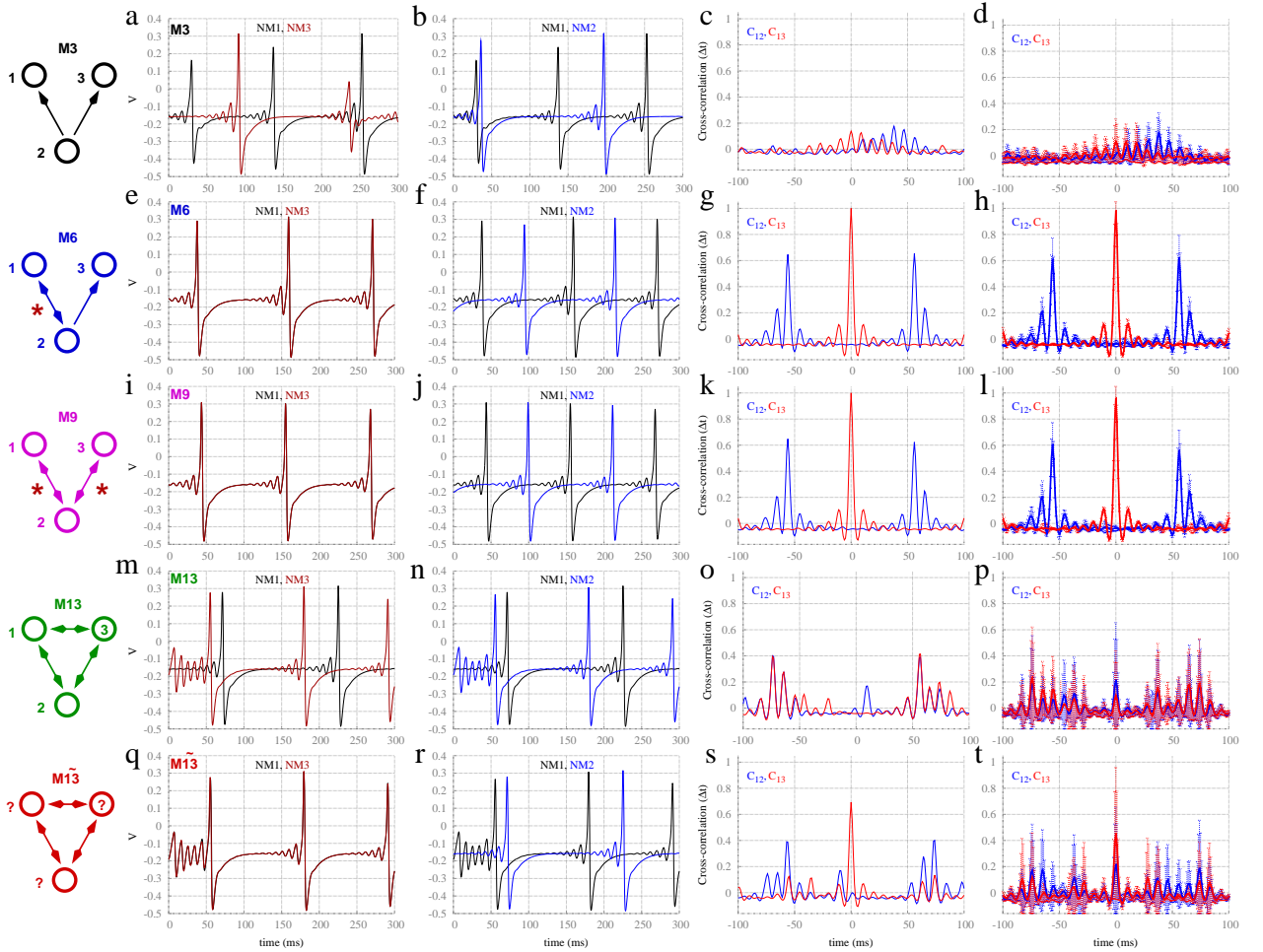}
\caption{\label{fig1}  
Synchronization in motifs of neural mass models. Dynamics of common driving motif M3 (a-d), common driving motifs with resonance pairs (tagged with red stars) M6 and M9 (e-l), and common driving motifs with frustrated resonant sources M13 and  M\~13 (m-t) with delay $\tau$ = 10 ms, and coupling strength c=0.01. 
 Motif M13 has node labels determined beforehand, whereas motif M\~13 has node labels determined \emph{a posteriori} such that the zero-lag synchronization between nodes  1 and 3 is maximized. First and second columns represent time traces of individual neural mass models (1 and 3,  1 and 2 respectively), whereas the third and forth columns represent the cross-correlation functions between node pairs of the corresponding single time series and averaged over 40 trials respectively with pairs 1-2 in blue and 1-3 in red.}
\end{center}
\end{figure}

We begin our analysis by focussing on 3-node motifs, whose enumeration (M1, M2, ..., M13) follows the seminal work of Sporns and K\"{o}tter (2004)~\cite{Sporns04}. As illustrated in the time traces for the common driving motif (M3) in the weak-coupling regime, robust synchronization between the driven nodes (1 and 3) does not occur (Fig. 1a), nor between directly coupled nodes (Fig. 1b). The lack of synchronization is also evident from the cross-correlation functions for one (Fig. 1 c) and for an average of 40 trials (Fig. 1 d). In contrast, strong zero-lag synchronization between these nodes (1 and 3) occurs in motifs (such as M6 and M9) that possess a reciprocal connection between at least one pair of nodes (Figs. 1 e-l). A crucial feature of the dynamics of these synchronized motifs (M6 and M9) is that whereas the pair of edge nodes 1 and 3 -- which are indirectly connected via node 2 -- synchronize in-phase, the pairs of neighbouring nodes (1-2 and 2-3) synchronize in anti-phase at the slow rhythm. 
Even with a coupling delay of 10 ms, owing to the internal dynamics of the neural mass models, the phase difference between the neighbouring nodes amounts to about 60 ms. Hence, mutually connected nodes enhance the synchronization in these small motifs, an effect that we previously showed that can propagate along chains of connected nodes~\cite{Gollo14}.

Importantly, this tendency of synchronized neighbouring nodes to oscillate in anti-phase cannot be satisfied for all motifs: For some configurations, for example when three nodes are mutually connected among themselves (motif M13), this tendency of synchronized neighbouring nodes to oscillate in antiphase is frustrated. Typically in this motif (Figs. 1 m-p) two out of the three pairs show anti-phase synchronization: When this happens, the third pair show in-phase synchronization. For example nodes 2 and 3 are in-phase for the case illustrated in Figs. 1 m-n. Hence, this corresponds to an example of frustrated dynamics because although each pair shows a tendency to oscillate in anti-phase, only two out of the three pairs can accomplish this anti-phase configuration at any one time.  

Starting from random initial conditions in this frustrated motif (M13), one cannot predict which pair of nodes will synchronize in-phase. Hence, the average zero-lag cross-correlation across trials between nodes 1 and 3 (Figs. 1 i-l), as well as any other pair of nodes (Fig. 1 p), is markedly diminished following the incorporation of the mutual connection between nodes 1 and 3. However, the selection of whichever two nodes have the maximum cross-correlation at zero-lag after each trial (and assigning them the labels 1 and 3 \emph{a posteriori}) reveals that the maximum zero-lag synchronization in this frustrated motif can also be strong: We called M\~{13} the auxiliary motif M13 that has nodes 1 and 3 labelled \emph{a posteriori} to maximize the zero-lag synchronization between them (Figs. 1 q-t).

\subsection*{Synchronization mode selection by symmetry breaking}

\begin{figure}[!h]
\begin{center}
\hspace{-0.5cm}
\includegraphics[angle=0,width=1.05\columnwidth]{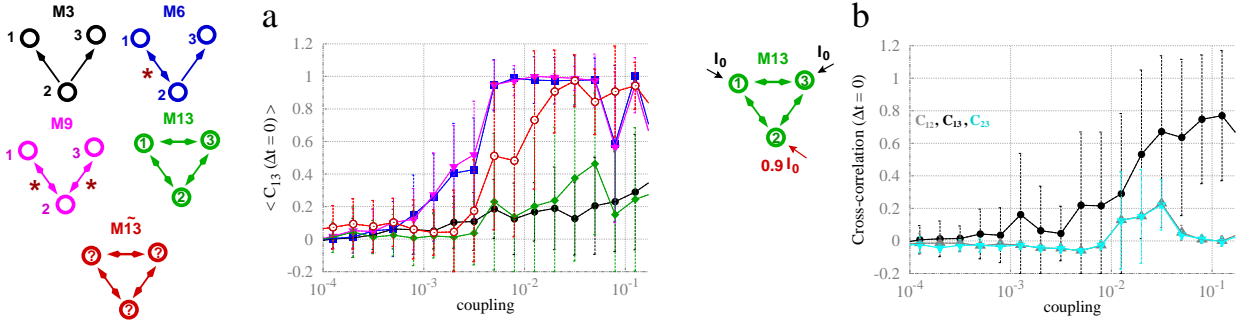}
\caption{\label{fig2}  
   Frustration and symmetry breaking. (a) Zero-lag crosscorrelation between neural
masses 1 and 3 for varying coupling strength. (b) Zero-lag crosscorrelation between pair of nodes for motif M13 with mismatch on the input current at node 2. 
 }
\end{center}
\end{figure}

The preceding analyses illustrate the effect that the resonance pair has on promoting synchronization and, conversely, how frustration destabilizes the stable state in weakly coupled small motifs. We next studied the robustness of these results with respect to the coupling strength (Fig. 2 a). Nodes 1 and 3 do not show in-phase synchrony for the common driving (M3) or frustrated (M13) motifs, but they synchronize even for very weak coupling for motifs which have resonance pairs (M6 and M9). Another motif that also exhibits zero-lag synchronization between nodes 1 and 3 is motif  M\~{13}. This shows that for sufficiently strong coupling, one pair of nodes synchronizes in phase in the frustrated motif. Despite the unpredictable character of the particular pair of nodes that synchronize when parameters across all nodes are identical, a small mismatch (such as a 10\% reduction in the input current over one neural mass model) breaks the symmetry, converting the distinct node to an apex. For example, a mismatch in 
node 2 favours phase synchronization between nodes 1-3 over the other pairs 1-2, and 2-3 (Fig. 2 b).

\subsection*{Synchronization on modular networks}

We now focus our analyses on canonical modular structures comprising two local communities connected through a single hub. In these prototypical modular networks, we set nodes to be mutually connected with every node within each module, and the hub is mutually connected to all other nodes. Such modular structures, even for the smallest cluster size of only two nodes per module (Fig. 3 a), contain frustrated motifs within clusters. In addition, this type of modular structure shares another common feature with small motifs: the hub plays the role of the apex node in motif M9, and the clusters mimic the outer nodes (Fig. 3 a).

 \begin{figure}[!h]
\begin{center}
\includegraphics[angle=0,width=0.99\columnwidth]{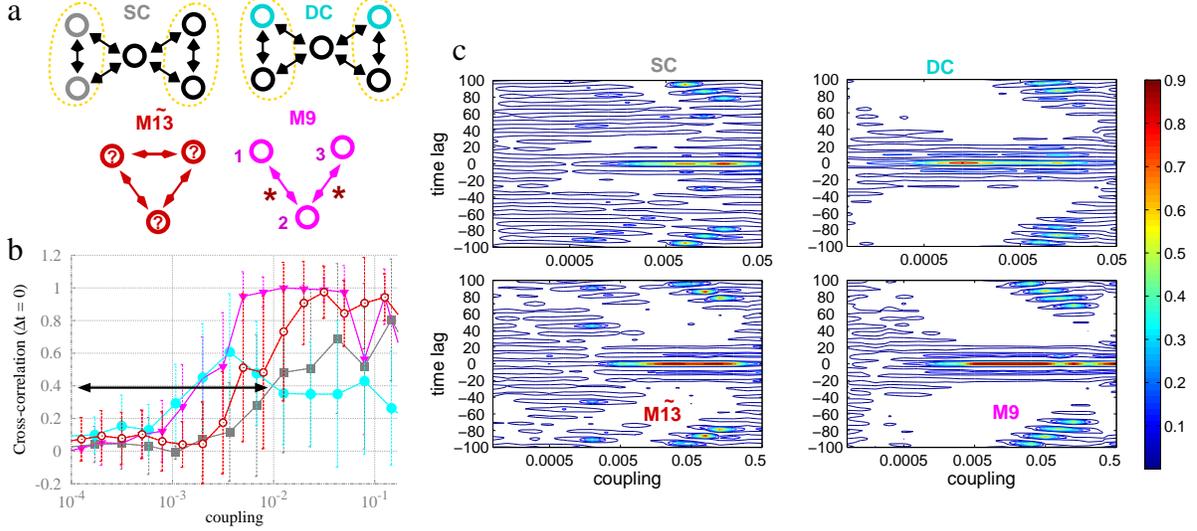}
\caption{\label{fig3}  
   Zero-lag synchronization from motifs to modular structures. (a) Pairs on minimal modular structure selected from the same cluster (SC) and from different clusters (DC) can be compared with motifs of 3-nodes. 
(b) Zero-lag synchronization of nodes 1 and 3 of motifs M9 and M\~13, and maximum zero-lag synchronization between nodes located at different and same clusters of modular structures.
Black arrow denotes the range of weak coupling in which the synchronization is stronger between
nodes of different clusters than nodes of the same cluster.
(c) Contour plot of crosscorrelations for varying coupling strength and time lag. Line represent curves along which the crosscorrelations have a constant value. White regions correspond to regions between contour lines, and the large white areas represent negative crosscorrelation values, as can be seen in Fig. 1k,l for motif M9 and c=0.01. Crosscorrelation between nodes 1 and 3 on motif M9 (M\~13) resembles the maximum
crosscorrelation between DC (SC) nodes.
 }
\end{center}
\end{figure}

To compare synchronization in modular structures against those in motifs, we focused on the cross-correlation between two nodes belonging either to the same cluster (SC), or to different clusters (DC). For each trial, the pair of nodes were selected \emph{a posteriori} as the two nodes exhibiting maximum zero-lag crosscorrelation. This was then averaged over all trials. The maximum phase synchronization between nodes belonging to different clusters resembles the phase synchronization between nodes 1 and 3 in motif M9, whereas the maximum  phase synchronization between nodes belonging to the same cluster resembles motif M\~13 (Figs. 3 b and c). This confirms that the motif structure is indeed influential in shaping these modular dynamics.

In the weak coupling regime, the maximum synchronization between different clusters is remarkably similar to the synchronization between nodes 1 and 3 for motif M9 (Fig. 3 b). However, for stronger coupling, when the synchronization within clusters grows, the synchronization between different clusters decays. 
As a result, the synchronization between clusters is only stronger than local synchronization for very weak coupling (see black arrow in Fig. 3 b).

We next tested the role of the intra-cluster connection strength in shaping the long-distance synchronization between clusters. This was achieved by multiplying the coupling strength of the intra-cluster connection $c$ by a weight factor w$\le 1$. For this experiment, we used a modular structure with a larger cluster size, as shown in Fig. 4 a. As illustrated in Figs. 4 b and c, the synchronization strength increases for both SC and DC as the weight w is reduced. However, although our measures of synchronization improve for very low w, the variability of the synchrony pattern is reduced because the frustration is diminished. Hence, in the case of weak intra-cluster connectivity, the system exhibits a strong tendency to evolve to a state of global synchronization in which all cluster nodes are in phase with each other, and in anti-phase synchrony with the hub node. 

 \begin{figure}[!h]
\begin{center}
\includegraphics[angle=0,width=0.99\columnwidth]{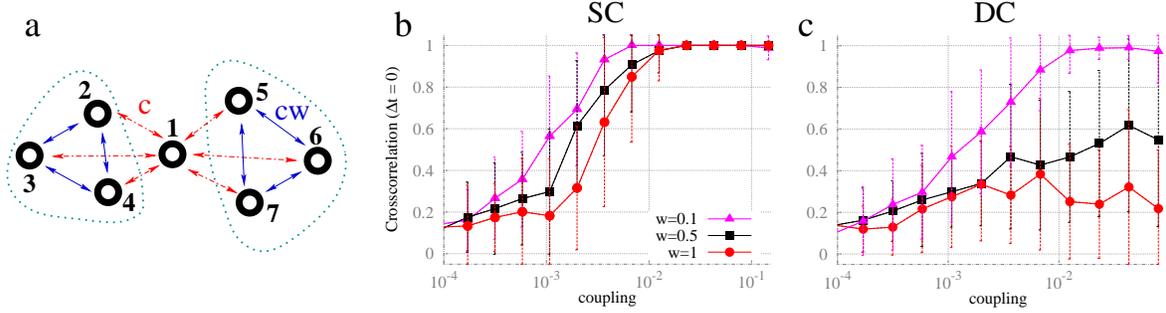}
\caption{\label{fig4}  
Maximum zero-lag synchronization in larger modular structures. (a) Canonical modular network with one hub (node 1) and two clusters (composed of nodes: 2, 3 and 4; 5, 6 and 7). The connections between hub-cluster nodes (red) have strength c, and cluster-cluster connections (blue) have strength cw. (b) Maximum zero-lag synchronization between nodes from the same cluster. (c) Maximum zero-lag synchronization between nodes located at different clusters. 
 }
\end{center}
\end{figure}

Setting the weight ratio w=1 and comparing the maximum zero-lag synchronization in modular structures with different cluster sizes (grey curve of Fig. 3 b against red curve of Fig. 4 b), we find that the synchronization inside a cluster is stronger for larger cluster sizes. Having the maximum synchronization inside a cluster that grows with the size is an expected result because for larger clusters the total coupling is stronger by construction (nodes inside a cluster are all-to-all connected), and the number of candidate pairs to synchronize also grows. On the other hand, an unexpected result appears when comparing the synchronization between clusters for different cluster sizes (blue curve of Fig. 3 b against red curve of Fig. 4 c): In this case, the larger cluster gives rise to a weaker synchronization between clusters. These results indicate that for w=1, there is a competition between SC and DC synchronization, in which the presence of SC synchronization prevents the synchronization between different 
clusters. 

\subsection*{Variability of synchronization patterns}

Even frustrated motifs show a small set of stable patterns of synchronization.  
Owing to the larger number of nodes and the presence of several frustrated motifs within each cluster, modular structures exhibit a larger number of stable patterns of synchronization. By performing multiple trials with random initial conditions, it is possible to estimate the variability of phase synchrony by counting the number of different outcomes that are found in a quasi-stationary regime. 

To measure the variability of phase synchrony, we dichotomised pairs as either synchronized or not synchronized for each trial of 2 sec (discarding a transient period of 500 ms) depending on whether the cross-correlation coefficient was greater than a threshold $\theta$. Figure 5 a illustrates different patterns of zero-lag synchronization obtained from the network studied in Fig. 4 a for different coupling strengths, in which white (black) stripes represent synchronized (unsynchronized) pairs. We then estimated the variability by counting the number of different patterns and dividing by 500, the total number of trials.

 \begin{figure}[!h]
\begin{center}
\includegraphics[angle=0,width=1.1\columnwidth]{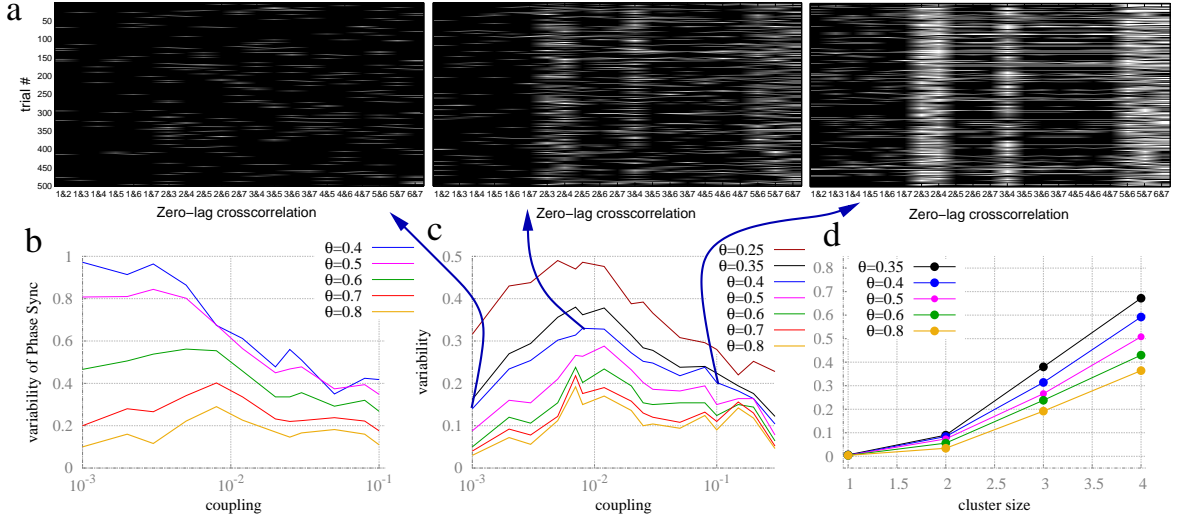}
\caption{\label{fig5}  
 Quantifying the variability of synchronization patterns in modular structures.
(a) Zero-lag crosscorrelation above the threshold $\theta$ between pairs of nodes (labelled as per Fig. 4 a)
in 500 trials. (b) Variability of the trial-to-trial patterns of synchronization for any phase lag in modular structures with cluster size 3.
(c) Same as panel b, but restricted to zero-lag synchronization.
(d) Variability of the zero-lag synchronization for modular structures with cluster of different sizes
and coupling $c=0.008$.
 }
\end{center}
\end{figure}

The analysis of the phase synchronization -- defined as the maximum cross-correlation at any lag (Fig. 5 b) -- reveals a strong dependence of the synchronization variability on the threshold $\theta$. The appearance of a maximum synchronization variability at intermediary coupling strength is evident for sufficiently high threshold. Next, we restricted the synchronization to zero-lag phase synchronization, such that a synchronized pair is identified only if the zero-lag cross-correlation coefficient was above $\theta$ (Fig. 5 c). Hence we see that the zero-lag synchronization variability is also optimized for an intermediary coupling strength: The synchronization is too sparse for very weak coupling and conversely too stable for strong coupling. Moreover, since the number of nodes, resonance pairs, and frustrated motifs grow with size, the maximal variability also grows with the cluster size (Fig. 5 d).

\section*{Macaque cortical networks}

\subsection*{Statistical features of Macaque structural networks}

   \begin{figure}[!h]
\begin{center}
\includegraphics[angle=0,width=0.95\columnwidth]{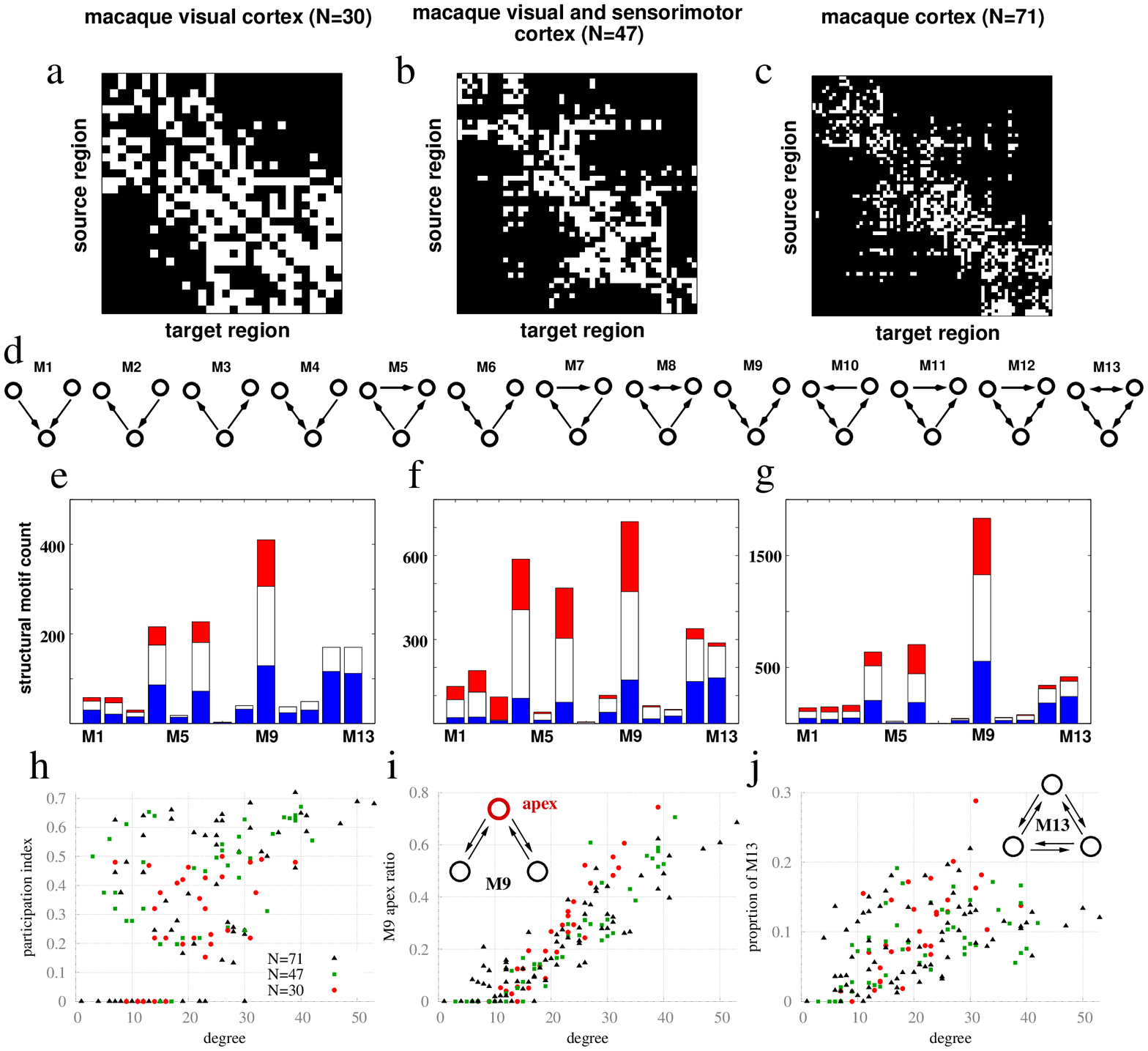}
\caption{\label{fig6}  
Motif distribution and modularity in macaque cortical networks. (a-c) Connectivity matrix for the different CoCoMac networks. White (black) squares indicate the presence (absence) of connections in these binary networks.  (d) 3-node motifs named as per Sporns and Kotter 2004. (e-g) Structural motif counts (SMC) for each cortical network. The lengths of the bars denote all SMC; blue partitions denote SMC with only intra-cluster connections; red partitions denote SMC with only inter-cluster connections; white partitions denote SMC with both intra- and inter-cluster connections. (h) Scatter plot of participation index (see appendix) and degree for each region of the three macaque networks. The degree is the sum of the number of inward and outward links. (i) Scatter plot of apex ratio for motif M9 and degree of each region. The apex ratio indicates the incidence of each region occupying the apex position of all motifs M9 that the region takes part. Hubs (nodes with high degree) exhibit high apex ration for motif M9. (j) Scatter plot of the proportion of counts of motif M13 that the regions participates in and degree of each region. 
 }
\end{center}
\end{figure}

We first re-visit some basic network features, such as modularity and motif distribution, as expressed in cortical networks of the macaque. These will guide our structure-function analyses. In particular we study three binary CoCoMac networks: The visual cortex (with N=30 nodes; Fig. 6a), the visual and sensorimotor areas (with N=47 nodes; Fig. 6 b), and the larger macaque cortical network (composed of N=71 nodes; Fig. 6 c). These are collated cortical networks of the macaque  (CoCoMac~\cite{Stephan01, Kotter04}), available at the brain connectivity toolbox~\cite{Rubinov10}.

A modular decomposition of these networks~\cite{Reichardt06,Leicht08, Rubinov10} suggests that these three networks have two, four, and four modules respectively. Analyses of the frequency of structural counts of the 13 configurations of 3-node motifs (Fig. 6d) for the macaque networks reveal substantial differences with respect to the presence of inter-modular connectivity (Figs. 6e-g).  Motifs M12 and M13 are typically composed entirely  by intra-modular connections (blue bars). In contrast, motifs M4, M6 and M9 appear more often connecting inter-modular nodes, either by exclusively inter-modular connections (red bars) or by a combination of inter- and intra-modular connections (white bars). The participation index (see appendix), which quantifies the inter- to intra-modular connectivity ratio of each node~\cite{Guimera05,Sporns07,Rubinov10}, shows that many nodes with low degree exhibit strictly intra-modular connections (null participation index; lower left hand corner of panel 6h). Conversely, high degree nodes typically interconnect modules (upper right hand corner). Between these extremes there exists a tremendous variety, such that nodes of intermediate degree may connect within and/or between modules. Interestingly, network motifs also show some degree-dependent statistical patterns in these macaque networks. High degree nodes, for example, tend to occupy the apex position of motif M9  (Fig. 6i)~\cite{Sporns07}. This feature suggests that nodes in the apex position of this bi-directionally connected 3-node chain may play an influential role on the dynamics of distant segregated areas. Additionally, excluding nodes with very low degree, the prototypical frustrated motif M13 appears to be expressed roughly independent of the node degree (Fig. 6j). This indicates that although this frustrated motif appears more often inside clusters, frustration can be a general feature affecting both hubs as well as peripheral nodes.

\subsection*{Synchronization patterns on Macaque cortex}

We finally extend our analyses of the variability of synchronization patterns to these macaque cortical networks. Consistent with the dynamics on our prototypical modular networks (Fig. 5 b), the variability of network patterns of zero-lag synchronization for the visual cortical network also exhibits a maximum for a moderate coupling strength (Fig. 7 a). Whilst our results so far were restricted to a delay coupling of 10 ms, it is important to note that the coupling delay also has a strong effect in the synchronization of the network and its variability (Fig. 7 b): Maximum variability appears for short ($\leq$10 ms) and very long ($>$40 ms) delays, with a minimum for intermediate values (15-20 ms). Also consistent with our prototypical networks, the dynamics on the larger cortical networks exhibit a richer variability of patterns of zero-lag synchronization  (Fig. 7 c). 

   \begin{figure}[!h]
\begin{center}
\includegraphics[angle=0,width=1.1\columnwidth]{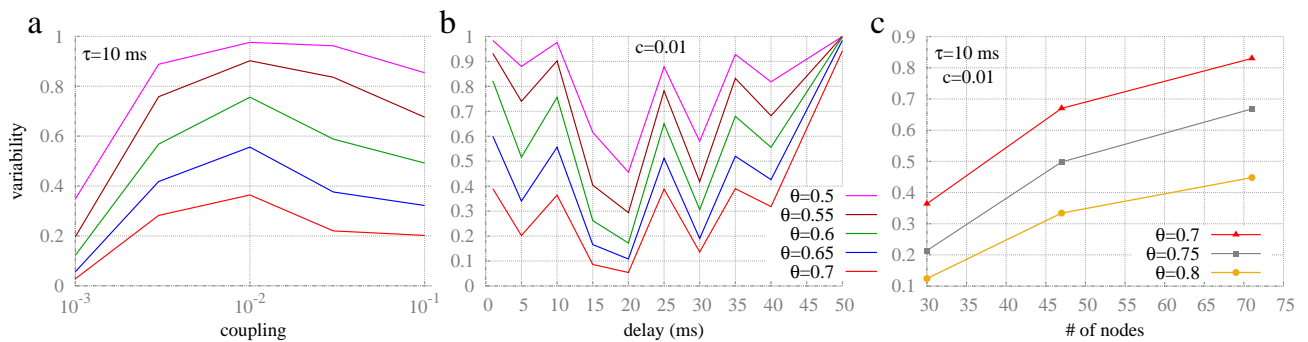}
\caption{\label{fig7}  
  Variability of synchronization patterns in cortical networks. Maximum variability of zero-lag synchronization, as measured in Fig. 5 c, for the macaque visual cortex (N=30) for varying coupling strength (a), and for varying coupling delay (b).  (c) Variability of zero-lag synchronization for different macaque cortical networks.
 }
\end{center}
\end{figure}

\section*{Discussion}

Adaptive brain function rests upon the ability of cortical networks to balance the competing constraints of variability and stability. Utilizing the dynamics on network motifs, we investigated the influence of structural connectivity on the diversity of synchronization patterns. We identified resonance and frustration as two major principles in sculpting the variability of the functional networks:  Reciprocal pairs enhance zero-lag synchronization via resonance-induced synchronization; Frustration decomposes stable synchronised dynamics into multiple competing stable solutions. Together, the two principles appear to give rise to metastable dynamics in systems of different sizes, from motifs to cortical networks. 

We propose that in larger networks, these dyadic influences reflect the contrasting deployment of these motifs with respects to their modular organization. Assuming that modules contain larger clustering, motif M13 represents a key connectivity pattern within modules. An important contribution of this frustrated motif is to enrich the dynamical landscape of synchronized cortical states. This property of frustrated systems to enrich and diversify the dynamics is general and occurs in other systems~\cite{levnajic11,jensen09,Kaluza10,Nisoli13,Nixon13}. However, to the best of our knowledge, this is the first time it has been related to large-scale cortical dynamics. 

The over-representation of motifs M9, and M6 is a consistent and characteristic property of cortical networks~\cite{Sporns04, Sporns07,Gollo14}. Taking into account that apex regions typically connect distant regions, in agreement with previous work~\cite{Vicente08,Gollo10,Gollo11}, our results suggest that one functional role for these hubs is to promote the synchronization between segregated regions belonging to different clusters. According to this proposition, these connections would enrich the variability of synchronization between clusters, which in our model occurs more sparsely than inside clusters.  

Changes in brain connectivity have been associated with pathological brain states, such as schizophrenia~\cite{Zalesky11} and multiple sclerosis~\cite{Coombs04}. Our results show that two crucial features of the structural connectivity that shape the cortical dynamics are the number of resonant pairs and frustrated motifs. Future work could investigate whether these two structural elements (resonant pairs and frustrated motifs) are altered in neurological and psychiatric illnesses characterized by “network dysplasia”~\cite{Bullmore97}. 

In addition to changes in structural connectivity, many pathological brain states are also associated with alterations in neuronal synchrony, which may be either reduced (e.g. autism~\cite{Wilson07} and schizophrenia~\cite{Breakspear03b,Spencer03}) or increased in levels of synchrony (e.g. epileptic seizures~\cite{Steriade03} and Parkinson's disease~\cite{Uhlhaas06b}). A balanced amount of synchronization is an essential feature of healthy neuronal dynamics~\cite{Friston00,Breakspear02b,Friston12}. Consistent with experiments and simulations in networks of spiking neurons~\cite{Yang12}, our numerical analyses of large-scale oscillatory cortical (modular and macaque) networks show that the maximum variability of synchronization patterns occurs for moderate coupling: Weaker coupling reduces the variability because fewer pairs can synchronize and the set of possible outcomes is restricted; stronger coupling reduces the variability because the networks often become globally synchronized, and a smaller set of stable 
states is favoured over others. Such changes might plausibly underlie the reduction in consciousness associated with anaesthesia~\cite{Murphy11}. 
This finding of maximal variability for moderate coupling reinforces the importance of homeostatic regulatory mechanisms to maintain the brain close to its optimal region of the parameter space~\cite{Friston12}.
Our results are also consistent with prior analyses of the role of delays~\cite{Deco09,Kutchko13} and network size in shaping the cortical dynamics.

Whilst resonance-induced synchronization has been shown to hold for synchronised pairs, regardless of their phase relation~\cite{Gollo14}, frustration requires the synchronization between pair of nodes to be in anti-phase.
Anti-phase synchronization is a common phase relationship between separated cortical regions found in a variety of experimental sets~\cite{Yang12,Canolty10}, and models~\cite{Li11,Vicente08,Gollo14}. 
The phase relation of the synchronization depends on many factors such as the balance between excitation and inhibition within a cortical region~\cite{Yang12}, and the coupling strength between regions~\cite{Li11}. 
More importantly, however, anti-phase becomes the dominant regime when inter-region synapses involve a substantial conduction delay~\cite{Li11}. 
For this reason we argue that our model, which favours anti-phase synchronization, is a suitable model for the purpose of understanding the dynamics in large-scale cortical networks, in which long distance connections are invariably associated with conduction latencies~\cite{Tomasi12}. Extension of our conclusions to other systems should fundamentally depend on whether anti-phase synchronization, the condition for frustration to occur, is the stable solution for two coupled nodes.

We have used a simple way of estimating the complexity of the landscape of cortical activity states. The advantage of this measure is that it is clear and intuitive to count the number of different possible dynamic outcomes. Future work could focus on more sophisticated measures of dynamic variability, based on the system entropy~\cite{Deco12}, phase reconfiguration~\cite{Breakspear04b} and the temporal dwelling in the various stable solutions~\cite{Freyer12}. Links to empirical measures of dynamic functional connectivity should also be forged~\cite{Zalesky14}. In addition, whilst we here forge a link between motifs and modules when bridging scales, future work could focus more closely on these related larger scales (of modules) as well as the shaping influence of the largest structural backbonme in the primate cortex, namely the rich club~\cite{van11}.

\subsection*{Appendix}

\subsubsection*{Neural mass models}

 The conductance-based neural mass model we used represents a mesoscale cortical region~\cite{Breakspear02}, which is derived from the biophysical Morris-Lecar model~\cite{Morris81}. 
 After adaptating to connect areas~\cite{Larter99} with synaptic interactions~\cite{Breakspear03}, it reached its most recent formulation, which is suitable to  model whole brain activity in large-scale networks~\cite{Honey07, Honey09, Gollo14}. 
Each cortical region was described by the neural mass model of spontaneous cortical dynamics:  
\begin{eqnarray}
\label{eqV}
\frac{dV^i (t)}{dt}&=& - \lbrace  g_{Ca}+ r_{NMDA} \,  a_{ee} \, [ (1-c) \,  Q^i_V(t)  \\ \nonumber
& & + c  \,  \langle Q^j_V (t-\tau)\rangle ] \rbrace   \,  m_{Ca}  \, (V^i(t)-V_{Ca}) \\ \nonumber
& & - \lbrace  g_{Na} \,  m_{Na}+ a_{ee} [(1-c) \,  Q^i_V(t)+ c  \,  \langle Q^j_V (t-\tau)\rangle ] \rbrace \,  (V^i(t)-V_{Na}) \\ \nonumber
& & -g_K \,  W^i(t)  \, (V^i(t)-V_K)- g_L \,  (V^i(t)-V_L) \\ \nonumber 
& & -a_{ie} \,  Z^i(t) \,  Q^i_Z+a_{ne}  \, I_0 \; , \\
\frac{dZ^i (t)}{dt}&=& b  \, \big( a_{ni} \,  I_0 + a_{ei} \,  V(t) \, Q^i_V(t) \big) \; ,\\
 \frac{d W^i(t)}{dt} &=&\frac{ \phi \, [m_K-W^i(t)]}{\tau_W}\;.
\end{eqnarray}
In the above equations $V$ is the mean membrane potential of the excitatory pyramidal neurons, $Z$ is the mean  membrane potential of the inhibitory interneurons and $W$ is the average number of open potassium ion channels. The neural-activation functions exhibit a sigmoidal-saturating grow with $V$ that governs the fraction of open channels $m_{ion}$: 
\begin{equation}
m_{ion}=0.5 \, \Bigg[1+\tanh \Bigg(\frac{V^i(t)-T_{ion}}{\delta_{ion}} \Bigg) \Bigg] \;.
\end{equation}
Finally, $Q^i_V$, and $Q^i_Z$ are the average neuronal firing rate of excitatory and inhibitory subpopulations of region $i$. 
Assuming Gaussian distributions, they are described by the following sigmoidal activation functions: 
\begin{eqnarray}
 Q^i_V(t) &=&0.5 \; Q_{V max} \;  \Bigg[1+\tanh \Bigg(\frac{V^i(t)-V_T}{\delta_{V}} \Bigg) \Bigg] \;, \\  
 Q^i_Z(t) &=&0.5 \; Q_{Z max}  \; \Bigg[1+\tanh \Bigg(\frac{Z^i(t)-Z_T}{\delta_{Z}} \Bigg) \Bigg] \;. 
\end{eqnarray}
In the above equations, the following set of parameters was used ~\cite{Breakspear03}: 
$g_{Ca}$(=1.1), $r_{NMDA}$(=0.25), $a_{ee}$(=0.4), $V_{Ca}$(=1), 
$g_{Na}$(=6.7), $V_{Na}$(=0.53), $g_K$(=2), $V_K$(=-0.7), 
$g_L$(=0.5), $V_L$(=-0.5), $a_{ie}$(=2), $a_{ne}$(=1), 
$I_0$(=0.3), $b$(=0.1), $a_{ni}$(=0.4), $a_{ei}$(=2), 
$T_{Ca}$(=-0.01), $T_{Na}$(=0.3), $T_{K}$(=0), 
$\delta_{Ca}$(=0.15), $\delta_{Na}$(=0.15), $\delta_{K}$(=0.3), 
$\phi$(=0.7), $\tau_W$(=1), 
$Q_{V max}$(=1), $V_T$(=0), $\delta_{V}$(=0.65), $Q_{Z max}$(=1), $Z_T$(=0), $\delta_{Z}$(=0.65).  
The above equations also include other varying parameters:  
the coupling strength between cortical regions $c$, 
the synaptic delay between cortical regions $\tau$, 
and $j$($=1, \, ... , \, N$), which are the afferent regions of region $i$. 
Consistent with the average values reported in macaques~\cite{Tomasi12},
we typically used a constant delay of 10 ms.
Although axonal delays between two given regions are typically distributed 
and specific roles have been attributed to non-homogeneous delays 
(e.g., to shape the EEG power spectrum~\cite{Roberts08}),
as a first approximation here we consider only homogeneous delays.
Small delay mismatches have been shown to not qualitatively affect the synchronization in neuronal motifs~\cite{Gollo14},
but the functions of delay diversity and distributed delays remain to be investigated in large-scale network models.  
The model was simulated in Matlab (Math Works) using the function {\it dde23}. 

\subsubsection*{Participation index}

The participation index $P$ quantifies how connected nodes are with regions belonging to other modules~\cite{Guimera05,Sporns07,Rubinov10}. 
It is defined as
\begin{equation}
P_{i}=1- \sum_{m=1}^{N_M} \left( \frac{\kappa_{im}}{k_i} \right)^2  \;,
\end{equation}
where the summation  $m$ runs over all identified modules $N_M$; $k_i$ and $\kappa_{im}$ stand for the degree of node $i$ and the number of edges from node $i$ to nodes within module $m$, respectively.

\section*{Acknowledgments}
We are thankful to Olaf Sporns, Claudio Mirasso, James Roberts, Andrew Zalesky, and  Chris Honey for valuable discussions.



\begin{thebibliography}{10}
\providecommand{\url}[1]{\texttt{#1}}
\providecommand{\urlprefix}{URL }
\expandafter\ifx\csname urlstyle\endcsname\relax
  \providecommand{\doi}[1]{doi:\discretionary{}{}{}#1}\else
  \providecommand{\doi}{doi:\discretionary{}{}{}\begingroup
  \urlstyle{rm}\Url}\fi
\providecommand{\bibAnnoteFile}[1]{%
  \IfFileExists{#1}{\begin{quotation}\noindent\textsc{Key:} #1\\
  \textsc{Annotation:}\ \input{#1}\end{quotation}}{}}
\providecommand{\bibAnnote}[2]{%
  \begin{quotation}\noindent\textsc{Key:} #1\\
  \textsc{Annotation:}\ #2\end{quotation}}
\providecommand{\eprint}[2][]{\url{#2}}

\bibitem{Passingham02}
Passingham RE, Stephan KE, K{\"o}tter R (2002) The anatomical basis of
  functional localization in the cortex.
\newblock Nature Reviews Neuroscience 3: 606--616.
\input{Passingham02}

\bibitem{Sporns10}
Sporns O (2010) Networks of the Brain, volume~1.
\newblock MIT Press, 375 pp.
\input{Sporns10}

\bibitem{Honey10}
Honey CJ, Thivierge JP, Sporns O (2010) Can structure predict function in the
  human brain?
\newblock Neuroimage 52: 766--776.
\input{Honey10}

\bibitem{Masquelier09}
Masquelier T, Hugues E, Deco G, Thorpe SJ (2009) Oscillations, phase-of-firing
  coding, and spike timing-dependent plasticity: an efficient learning scheme.
\newblock Journal of Neuroscience 29: 13484--13493.
\input{Masquelier09}

\bibitem{Fries05}
Fries P (2005) A mechanism for cognitive dynamics: neuronal communication
  through neuronal coherence.
\newblock Trends in Cognitive Sciences 9: 474-480.
\input{Fries05}

\bibitem{Breakspear04}
Breakspear M (2004) {"Dynamic"} connectivity in neural systems: theoretical and
  empirical considerations.
\newblock Neuroinformatics 2: 205.
\input{Breakspear04}

\bibitem{Hutchison13}
Hutchison RM, Womelsdorf T, Allen EA, Bandettini PA, Calhoun VD, et~al. (2013)
  Dynamic functional connectivity: Promises, issues, and interpretations.
\newblock NeuroImage 80: 360-78.
\input{Hutchison13}


\bibitem{Zalesky14}
Zalesky A, Fornito A, Cocchi L, Gollo LL, Breakspear M. 2014 Time-resolved resting-state brain networks. 
\newblock Proceedings of the National Academy of Sciences, 111(28), 10341--10346.
\input{Zalesky14}

\bibitem{Friston00}
Friston KJ (2000) The labile brain. i. neuronal transients and nonlinear
  coupling.
\newblock Philosophical Transactions of the Royal Society of London Series B:
  Biological Sciences 355: 215--236.
\input{Friston00}

\bibitem{Breakspear02b}
Breakspear M (2002) Nonlinear phase desynchronization in human
  electroencephalographic data.
\newblock Human brain mapping 15: 175--198.
\input{Breakspear02b}

\bibitem{Friston12}
Friston K, Breakspear M, Deco G (2012) Perception and self-organized
  instability.
\newblock Frontiers in computational neuroscience 6:44.
\input{Friston12}

\bibitem{Kutchko13}
Kutchko KM, Fr{\"o}hlich F (2013) Emergence of metastable state dynamics in
  interconnected cortical networks with propagation delays.
\newblock PLoS computational biology 9: e1003304.
\input{Kutchko13}

\bibitem{Yang12}
Yang H, Shew WL, Roy R, Plenz D (2012) Maximal variability of phase synchrony
  in cortical networks with neuronal avalanches.
\newblock The Journal of Neuroscience 32: 1061--1072.
\input{Yang12}

\bibitem{Milo02}
Milo R, Shen-Orr S, Itzkovitz S, Kashtan N, Chklovskii D, et~al. (2002) Simple
  building blocks of complex networks.
\newblock Science 298: 824 - 827.
\input{Milo02}

\bibitem{Sporns04}
Sporns O, K\"{o}tter R (2004) Motifs in brain networks.
\newblock PLoS Biology 2: e369.
\input{Sporns04}

\bibitem{Song05}
Song S, Sj{\"o}str{\"o}m PJ, Reigl M, Nelson S, Chklovskii DB (2005) Highly
  nonrandom features of synaptic connectivity in local cortical circuits.
\newblock PLoS biology 3: e68.
\input{Song05}

\bibitem{Sporns07}
Sporns O, Honey CJ, K{\"o}tter R (2007) Identification and classification of
  hubs in brain networks.
\newblock PloS one 2: e1049.
\input{Sporns07}

\bibitem{Rubinov10}
Rubinov M, Sporns O (2010) Complex network measures of brain connectivity: uses
  and interpretations.
\newblock Neuroimage 52: 1059--1069.
\input{Rubinov10}

\bibitem{Harriger12}
Harriger L, van~den Heuvel MP, Sporns O (2012) Rich club organization of
  macaque cerebral cortex and its role in network communication.
\newblock PloS one 7: e46497.
\input{Harriger12}

\bibitem{Eguiluz11}
Egu{\'\i}luz VM, P{\'e}rez T, Borge-Holthoefer J, Arenas A (2011) Structural
  and functional networks in complex systems with delay.
\newblock Physical Review E 83: 056113.
\input{Eguiluz11}

\bibitem{Zhao11}
Zhao L, Bryce~Beverlin I, Netoff T, Nykamp DQ (2011) Synchronization from
  second order network connectivity statistics.
\newblock Frontiers in Computational Neuroscience 5: 1-16.
\input{Zhao11}

\bibitem{Trousdale12}
Trousdale J, Hu Y, Shea-Brown E, Josi{\'c} K (2012) Impact of network structure
  and cellular response on spike time correlations.
\newblock PLoS computational biology 8: e1002408.
\input{Trousdale12}

\bibitem{Gollo14}
Gollo LL, Mirasso C, Sporns O, Breakspear M (2013) Mechanisms of zero-lag
  synchronization in cortical motifs.
\newblock arXiv preprint arXiv:13045008 .
\input{Gollo14}

\bibitem{Bullmore09}
Bullmore E, Sporns O (2009) Complex brain networks: graph theoretical analysis
  of structural and functional systems.
\newblock Nature Reviews Neuroscience 10: 186--198.
\input{Bullmore09}

\bibitem{Felleman91}
Felleman DJ, Van~Essen DC (1991) Distributed hierarchical processing in the
  primate cerebral cortex.
\newblock Cerebral Cortex 1: 1--47.
\input{Felleman91}

\bibitem{Breakspear05}
Breakspear M, Stam CJ (2005) Dynamics of a neural system with a multiscale
  architecture.
\newblock Philosophical Transactions of the Royal Society of London - Series B:
  Biological Sciences 360: 1051--1074.
\input{Breakspear05}

\bibitem{Stam12}
Stam C, Van~Straaten E (2012) The organization of physiological brain networks.
\newblock Clinical Neurophysiology 123: 1067--1087.
\input{Stam12}

\bibitem{Hilgetag04}
Hilgetag CC, Kaiser M (2004) Clustered organization of cortical connectivity.
\newblock Neuroinformatics 2: 353--360.
\input{Hilgetag04}

\bibitem{Zemanova06}
Zemanov\'{a} L, Zhou C, Kurths J (2006) Structural and functional clusters of
  complex brain networks.
\newblock Physica D: Nonlinear Phenomena 224: 202--212.
\input{Zemanova06}

\bibitem{Hagmann08}
Hagmann P, Cammoun L, Gigandet X, Meuli R, Honey CJ, et~al. (2008) Mapping the
  structural core of human cerebral cortex.
\newblock PLoS Biol 6: e159.
\input{Hagmann08}

\bibitem{Zamora10}
Zamora-L{\'o}pez G, Zhou C, Kurths J (2010) Cortical hubs form a module for
  multisensory integration on top of the hierarchy of cortical networks.
\newblock Frontiers in neuroinformatics 4.
\input{Zamora10}

\bibitem{van13}
van~den Heuvel MP, Sporns O (2013) Network hubs in the human brain.
\newblock Trends in cognitive sciences 17: 683--696.
\input{van13}

\bibitem{Bullmore12}
Bullmore E, Sporns O (2012) The economy of brain network organization.
\newblock Nature Reviews Neuroscience 13: 336--349.
\input{Bullmore12}

\bibitem{Chen13}
Chen Y, Wang S, Hilgetag CC, Zhou C (2013) Trade-off between multiple
  constraints enables simultaneous formation of modules and hubs in neural
  systems.
\newblock PLoS computational biology 9: e1002937.
\input{Chen13}

\bibitem{Zhou06}
Zhou C, Zemanov\'a L, Zamora G, Hilgetag CC, Kurths J (2006) Hierarchical
  organization unveiled by functional connectivity in complex brain networks.
\newblock Phys Rev Lett 97: 238103.
\input{Zhou06}

\bibitem{Honey07}
Honey C, K\"{o}tter R, Breakspear M, Sporns O (2007) Network structure of
  cerebral cortex shapes functional connectivity on multiple time scales.
\newblock Proc Natl Acad Sci 104: 10240--10245.
\input{Honey07}

\bibitem{Zhou07}
Zhou C, Zemanov\'a L, Zamora-L\'{o}pez G, Hilgetag CC, Kurths J (2007)
  Structure--function relationship in complex brain networks expressed by
  hierarchical synchronization.
\newblock New Journal of Physics 9: 178--178.
\input{Zhou07}

\bibitem{Breakspear02}
Breakspear M, Terry JR (2002) Nonlinear interdependence in neural systems:
  motivation, theory, and relevance.
\newblock The International journal of neuroscience 112: 1263--1284.
\input{Breakspear02}

\bibitem{Honey09}
Honey CJ, Sporns O, Cammoun L, Gigandet X, Thiran JP, et~al. (2009) Predicting
  human resting-state functional connectivity from structural connectivity.
\newblock Proc Natl Acad Sci 106: 2035--2040.
\input{Honey09}

\bibitem{Tomasi12}
Tomasi S, Caminiti R, Innocenti GM (2012) Areal differences in diameter and
  length of corticofugal projections.
\newblock Cerebral Cortex 22: 1463--1472.
\input{Tomasi12}

\bibitem{Arieli96}
Arieli A, Sterkin A, Grinvald A, Aersten A (1996) Dynamics of ongoing activity:
  explanation of the large variability in evoked cortical responses.
\newblock Science 273: 1868-1871.
\input{Arieli96}

\bibitem{Poulet08}
Poulet JF, Petersen CC (2008) Internal brain state regulates membrane potential
  synchrony in barrel cortex of behaving mice.
\newblock Nature 454: 881--885.
\input{Poulet08}

\bibitem{Cafaro10}
Cafaro J, Rieke F (2010) Noise correlations improve response fidelity and
  stimulus encoding.
\newblock Nature 468: 964--967.
\input{Cafaro10}

\bibitem{Stephan01}
Stephan KE, Kamper L, Bozkurt A, Burns GA, Young MP, et~al. (2001) Advanced
  database methodology for the collation of connectivity data on the macaque
  brain (cocomac).
\newblock Philosophical Transactions of the Royal Society of London Series B:
  Biological Sciences 356: 1159--1186.
\input{Stephan01}

\bibitem{Kotter04}
K\"{o}tter R (2004) Online retrieval, processing, and visualization of primate
  connectivity data from the cocomac database.
\newblock Neuroinformatics 2: 127--144.
\input{Kotter04}

\bibitem{Reichardt06}
Reichardt J, Bornholdt S (2006) Statistical mechanics of community detection.
\newblock Physical Review E 74: 016110.
\input{Reichardt06}

\bibitem{Leicht08}
Leicht EA, Newman ME (2008) Community structure in directed networks.
\newblock Physical review letters 100: 118703.
\input{Leicht08}

\bibitem{Guimera05}
Guimera R, Amaral LAN (2005) Functional cartography of complex metabolic
  networks.
\newblock Nature 433: 895--900.
\input{Guimera05}

\bibitem{levnajic11}
Levnaji{\'c} Z (2011) Emergent multistability and frustration in
  phase-repulsive networks of oscillators.
\newblock Physical Review E 84: 016231.
\input{levnajic11}

\bibitem{jensen09}
Jensen MH, Krishna S, Pigolotti S (2009) Repressor lattice: feedback,
  commensurability, and dynamical frustration.
\newblock Physical review letters 103: 118101.
\input{jensen09}

\bibitem{Kaluza10}
Kaluza P, Meyer-Ortmanns H (2010) On the role of frustration in excitable
  systems.
\newblock Chaos: An Interdisciplinary Journal of Nonlinear Science 20:
  043111--043111.
\input{Kaluza10}

\bibitem{Nisoli13}
Nisoli C, Moessner R, Schiffer P (2013) Colloquium: Artificial spin ice:
  Designing and imaging magnetic frustration.
\newblock Reviews of Modern Physics 85: 1473.
\input{Nisoli13}

\bibitem{Nixon13}
Nixon M, Ronen E, Friesem AA, Davidson N (2013) Observing geometric frustration
  with thousands of coupled lasers.
\newblock Physical review letters 110: 184102.
\input{Nixon13}

\bibitem{Vicente08}
Vicente R, Gollo LL, Mirasso CR, Fischer I, Pipa G (2008) Dynamical relaying
  can yield zero time lag neuronal synchrony despite long conduction delays.
\newblock Proc Natl Acad Sci 105: 17157-17162.
\input{Vicente08}

\bibitem{Gollo10}
Gollo LL, Mirasso C, Villa AEP (2010) Dynamic control for synchronization of
  separated cortical areas through thalamic relay.
\newblock NeuroImage 52: 947--955.
\input{Gollo10}

\bibitem{Gollo11}
Gollo LL, Mirasso CR, Atienza M, Crespo-Garcia M, Cantero JL (2011) Theta band
  zero-lag long-range cortical synchronization via hippocampal dynamical
  relaying.
\newblock PLoS ONE 6: e17756.
\input{Gollo11}

\bibitem{Zalesky11}
Zalesky A, Fornito A, Seal ML, Cocchi L, Westin CF, et~al. (2011) Disrupted
  axonal fiber connectivity in schizophrenia.
\newblock Biological psychiatry 69: 80--89.
\input{Zalesky11}

\bibitem{Coombs04}
Coombs BD, Best A, Brown MS, Miller DE, Corboy J, et~al. (2004) Multiple
  sclerosis pathology in the normal and abnormal appearing white matter of the
  corpus callosum by diffusion tensor imaging.
\newblock Multiple sclerosis 10: 392--397.
\input{Coombs04}

\bibitem{Bullmore97}
Bullmore E, Frangou S, Murray R (1997) The dysplastic net hypothesis: an
  integration of developmental and dysconnectivity theories of schizophrenia.
\newblock Schizophrenia research 28: 143--156.
\input{Bullmore97}

\bibitem{Wilson07}
Wilson TW, Rojas DC, Reite ML, Teale PD, Rogers SJ (2007) Children and
  adolescents with autism exhibit reduced meg steady-state gamma responses.
\newblock Biological psychiatry 62: 192--197.
\input{Wilson07}

\bibitem{Breakspear03b}
Breakspear M, Terry J, Friston K, Harris A, Williams L, et~al. (2003) A
  disturbance of nonlinear interdependence in scalp {EEG} of subjects with
  first episode schizophrenia.
\newblock Neuroimage 20: 466--478.
\input{Breakspear03b}

\bibitem{Spencer03}
Spencer KM, Nestor PG, Niznikiewicz MA, Salisbury DF, Shenton ME, et~al. (2003)
  Abnormal neural synchrony in schizophrenia.
\newblock The Journal of Neuroscience 23: 7407--7411.
\input{Spencer03}

\bibitem{Steriade03}
Steriade M (2003) Neuronal substrates of sleep and epilepsy.
\newblock Cambridge University Press.
\input{Steriade03}

\bibitem{Uhlhaas06b}
Uhlhaas PJ, Singer W (2006) Neural synchrony in brain disorders: relevance for
  cognitive dysfunctions and pathophysiology.
\newblock Neuron 52: 155--168.
\input{Uhlhaas06b}

\bibitem{Murphy11}
Murphy M, Bruno MA, Riedner BA, Boveroux P, Noirhomme Q, et~al. (2011) Propofol
  anesthesia and sleep: a high-density {EEG} study.
\newblock Sleep 34: 283.
\input{Murphy11}

\bibitem{Deco09}
Deco G, Jirsa V, McIntosh A, Sporns O, K{\"o}tter R (2009) Key role of
  coupling, delay, and noise in resting brain fluctuations.
\newblock Proceedings of the National Academy of Sciences 106: 10302--10307.
\input{Deco09}

\bibitem{Canolty10}
Canolty RT, Ganguly K, Kennerley SW, Cadieu CF, Koepsell K, et~al. (2010)
  Oscillatory phase coupling coordinates anatomically dispersed functional cell
  assemblies.
\newblock Proc Natl Acad Sci 107: 17356--17361.
\input{Canolty10}

\bibitem{Li11}
Li D, Zhou C (2011) Organization of anti-phase synchronization pattern in
  neural networks: what are the key factors?
\newblock Frontiers in systems neuroscience 5.
\input{Li11}

\bibitem{Deco12}
Deco G, Jirsa VK (2012) Ongoing cortical activity at rest: criticality,
  multistability, and ghost attractors.
\newblock The Journal of Neuroscience 32: 3366--3375.
\input{Deco12}

\bibitem{Breakspear04b}
Breakspear M, Williams LM, Stam CJ (2004) A novel method for the topographic
  analysis of neural activity reveals formation and dissolution of ‘dynamic
  cell assemblies’.
\newblock Journal of Computational Neuroscience 16: 49--68.
\input{Breakspear04b}

\bibitem{Freyer12}
Freyer F, Roberts JA, Ritter P, Breakspear M (2012) A canonical model of
  multistability and scale-invariance in biological systems.
\newblock PLoS computational biology 8: e1002634.
\input{Freyer12}

\bibitem{van11}
van~den Heuvel MP, Sporns O (2011) Rich-club organization of the human
  connectome.
\newblock The Journal of neuroscience 31: 15775--15786.
\input{van11}

\bibitem{Morris81}
Morris C, Lecar H (1981) Voltage oscillations in the barnacle giant muscle
  fiber.
\newblock Biophys J 35: 193-213.
\input{Morris81}

\bibitem{Larter99}
Larter R, Speelman B, Worth RM (1999) A coupled ordinary differential equation
  lattice model for the simulation of epileptic seizures.
\newblock Chaos: An Interdisciplinary Journal of Nonlinear Science 9: 795-804.
\input{Larter99}

\bibitem{Breakspear03}
Breakspear M, Terry JR, Friston KJ (2003) Modulation of excitatory synaptic
  coupling facilitates synchronization and complex dynamics in a biophysical
  model of neuronal dynamics.
\newblock Network: Computation in Neural Systems 14: 703--732.
\input{Breakspear03}

\bibitem{Roberts08}
Roberts JA, Robinson PA (2008) Modeling distributed axonal delays in mean-field
  brain dynamics.
\newblock Phys Rev E 78: 051901.
\input{Roberts08}

\end{thebibliography}

\end{document}